\documentclass[conference]{IEEEtran}
\IEEEoverridecommandlockouts
\usepackage{amsmath,amssymb,amsfonts}
\usepackage{algorithmic}
\usepackage{graphicx}
\usepackage{textcomp}
\usepackage{xcolor}
\usepackage{cleveref}
\usepackage{subcaption}
\usepackage{makecell}
\usepackage{tikz}
\usepackage{pgfplots}

\usepackage[style=ieee]{biblatex}
\Crefname{figure}{Fig.}{Figs.}

\def\BibTeX{{\rm B\kern-.05em{\sc i\kern-.025em b}\kern-.08em
    T\kern-.1667em\lower.7ex\hbox{E}\kern-.125emX}}
\begin{document}

\title{GPU Acceleration of 3D Agent-Based\\Biological Simulations}

\author{\IEEEauthorblockN{Ahmad Hesam}
\IEEEauthorblockA{\textit{ABS group}\\
\textit{Delft University of Technology}\\
Delft, Netherlands\\
a.s.hesam@tudelft.nl}
\and
\IEEEauthorblockN{Lukas Breitwieser}
\IEEEauthorblockA{\textit{CERN openlab}\\
\textit{CERN}\\
Geneva, Switzerland\\
lukas.breitwieser@cern.ch}
\and
\IEEEauthorblockN{Fons Rademakers}
\IEEEauthorblockA{\textit{CERN openlab}\\
\textit{CERN}\\
Geneva, Switzerland\\
fons.rademakers@cern.ch}
\and
\IEEEauthorblockN{Zaid Al-Ars}
\IEEEauthorblockA{\textit{ABS group}\\
\textit{Delft University of Technology}\\
Delft, Netherlands\\
z.al-ars@tudelft.nl}
}

\maketitle

\begin{abstract}
Researchers in biology are faced with the tough challenge of developing high-performance computer simulations of their increasingly complex agent-based models.
BioDynaMo is an open-source agent-based simulation platform that aims to alleviate researchers from the intricacies that go into the development of high-performance computing.
Through a high-level interface, researchers can implement their models on top of BioDynaMo’s multi-threaded core execution engine to rapidly develop simulations that effectively utilize parallel computing hardware.
In biological agent-based modeling, the type of operations that are typically the most compute-intensive are those that involve agents interacting with their local neighborhood.
In this work, we investigate the currently implemented method of handling neighborhood interactions of cellular agents in BioDynaMo, and ways to improve the performance to enable large-scale and complex simulations.
We propose to replace the kd-tree implementation to find and iterate over the neighborhood of each agent with a uniform grid method that allows us to take advantage of the massively parallel architecture of graphics processing units (GPUs).
We implement the uniform grid method in both CUDA and OpenCL to address GPUs from all major vendors and evaluate several techniques to further improve the performance.
Furthermore, we analyze the performance of our implementations for models with a varying density of neighboring agents.
As a result, the performance of the mechanical interactions method improved by up to two orders of magnitude in comparison to the multithreaded baseline version.
The implementations are open-source and publicly available on Github.
\end{abstract}


\section{Introduction}
\par Agent-based simulation (ABS) is a powerful tool for conducting research on complex biological systems.
In ABS, a biological system is composed of a number of agents that individually are modeled to follow a fixed set of, often simple, rules.
Agents can interact with neighboring agents or respond to external stimuli.
Although the individual behavior of agents is often trivial, the emerging behavior that comes forth from the biological system as a whole can give researchers valuable insights \cite{macal2009agent, an2009agent, di2006vivo}.
\par As the complexity and scale of biological agent-based models increases so does the demand for computational power and efficiency \cite{an2009agent}.
Agent-based simulations are inherently parallelizable in their execution, as the agents' states can be modified independently of each other.
Modern-day hardware is becoming increasingly more parallelized as a result of Dennard scaling \cite{fiori2014electronics} and the stagnation of Moore's law \cite{moore1965cramming}, as pointed out in \cite{breitwieser2021biodynamo}.
Moreover, general-purpose computing on graphics processing units (GPUs) is an attractive solution to improve the computational efficiency of ABS applications in particular~\cite{lysenko2008, richmond_high_2010}, and parallel applications in general~\cite{gpu_genomics,gpu_brain}.
By porting applications to, either fully or partially, run on GPUs it is possible to observe speedups of several orders of magnitude in comparison to the CPU-only execution \cite{nickolls2010gpu}.
Although several ABS frameworks exist that achieve significant speedups using GPUs in the field of ABS, there is still significant room for improvement, which we wish to address in this article.
\par BioDynaMo \cite{breitwieser2021biodynamo} is an open-source software platform for life scientists for simulating biological agent-based models.
Each agent in BioDynamo is programmed to follow a specified set of rules, imposed by the modeler, that can trigger specified actions affecting itself or other agents.
Agents in biological systems often interact with their local environment, and their behavior can be influenced by other agents that reside within a certain range.
An example is the mechanical interactions a cellular agent undergoes when it physically collides with another agent.
Local interactions are an extremely important concept in biological systems since it is the driving force behind key biological processes, such as tissue development \cite{van2015simulating}.
\par BioDynaMo is fully parallelized using OpenMP and its performance scales with the number of CPU cores available on a system \cite{breitwieser2021biodynamo}.
To further enhance the simulations' performance, we want to investigate the applicability of GPUs in accelerating compute-intensive operations in BioDynaMo.
In this work we present the following contributions:
\let\labelitemi\labelitemii
\begin{itemize}
    \item Redesign the neighborhood search in BioDynaMo from a kd-tree method to a uniform grid method to profit from the parallel architecture of GPUs.
    \item Port the uniform grid implementation to GPU code using OpenCL and CUDA to address all major GPU vendors.
    \item Improve the GPU kernels based on domain-specific aspects of biological agent-based models.
    \item Benchmark the runtime and analyze the performance gains that are obtained.
\end{itemize}

\par The organization of the paper is as follows.
\Cref{sec:related_work} discusses related work.
In \Cref{sec:problem_definition} we define the problem in more detail.
In \Cref{sec:methodology} we describe the methodology of our approach.
\Cref{sec:experimental_setup} describes the hardware and software setup.
In \Cref{sec:results} we present the results.
And finally, in \Cref{sec:conclusions} we draw the conclusions of this work.

\section{Related Work}\label{sec:related_work}
\par There are several frameworks and software packages that make it possible to simulate agent-based models for biological systems.
There are many more specialized software solutions, but these generally focus on one biological process, or a few closely related biological processes.
Some of the more general ABS frameworks for biological systems (BioCellion \cite{kang2014biocellion}, PhysiCell \cite{ghaffarizadeh2018physicell}, Timothy \cite{cytowski2014large}, and Chaste \cite{mirams2013chaste}) 
focus, among other things, on computational efficiency, but do not support GPU acceleration. 
In this work, we demonstrate that GPU acceleration is possible for general-purpose agent-based platforms.
\par In the works of \cite{lysenko2008} and \cite{richmond_high_2010} the authors present cellular agent-based simulation (ABS) programs that run entirely on a GPU.
The authors report speedups of several orders of magnitude over ABS frameworks that are only targeted for CPUs.
Although the findings are impressive, the fact that the simulation runs entirely on the GPU has two major drawbacks.
First, it puts a lot of pressure on minimizing memory consumption.
As GPU memory is a non-expandable and limited resource, there is a limit to the complexity of the agents' state and the scale of the model.
In this work, we offload the most compute-intensive operation to GPU, which requires only a subset of the agents' state data to be present on the GPU memory.
Second, operations that are independent of the agents, such as extracellular substance diffusion, are integral to biological systems and are absent from these works.
With BioDynaMo we can simulate the extracellular substance diffusion efficiently on a multi-core CPU, independently from the GPU operations \cite{breitwieser2021biodynamo}.

\begin{figure}[t]
    \centering
    \includegraphics[width=0.3\textwidth]{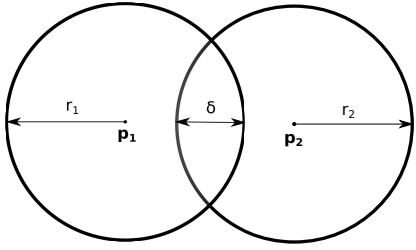}
    \caption{Sphere-sphere collision force diagram (projected as circles for simplicity).}
    \label{fig:default_force}
\end{figure}

\section{Problem Definition} \label{sec:problem_definition}
\par The mechanical interaction operation is one of the most compute-intensive operations in any cellular agent-based model.
Each cell (i.e. agent) interacts with all other cells within a certain interaction radius.
For cells that are physically in contact with each other, we need to compute the collision forces and the resulting displacement.
In BioDynaMo cellular agents can be physically modeled as spherical objects.
For the scope of this paper, we shall consider only sphere-sphere interactions, as illustrated in \Cref{fig:default_force} (projected as circles).
\Cref{eq:default_force} \cite{hauri2013self} shows the calculations involved in determining the mechanical force.

\begin{equation}
    \begin{array}{lcl}
        \delta & = & r_1 + r_2 - \lVert \mathbf{p_1} - \mathbf{p_2}\rVert \\[2pt]
        r & = & \frac{r_1 \cdot r_2}{r_1 + r_2} \\[2pt]
        \mathbf{F} & = & (\kappa \cdot \delta - \gamma \cdot \sqrt{r \cdot \delta}) \cdot \frac{\mathbf{p_1} - \mathbf{p_2}}{\lVert \mathbf{p_1} - \mathbf{p_2}\rVert}\,,
    \end{array}
    \label{eq:default_force}
\end{equation}
where $r_1$ and $r_2$ are the radii of the spheres, $\mathbf{p_1}$ and $\mathbf{p_2}$ their position vectors, $\kappa$ the repulsion coefficient, $\gamma$ the attraction coefficient, and $\mathbf{F}$ the resulting collision force vector.
After the collision force has been computed, we determine whether it is strong enough to break the adherence of the cell in question.
If that is the case, then we integrate over the collision force to compute the final displacement.
The length of the final displacement vector is generally limited by an upper bound.
\par
\begin{figure}[!t]
    \centering
    \includegraphics[width=0.4\textwidth]{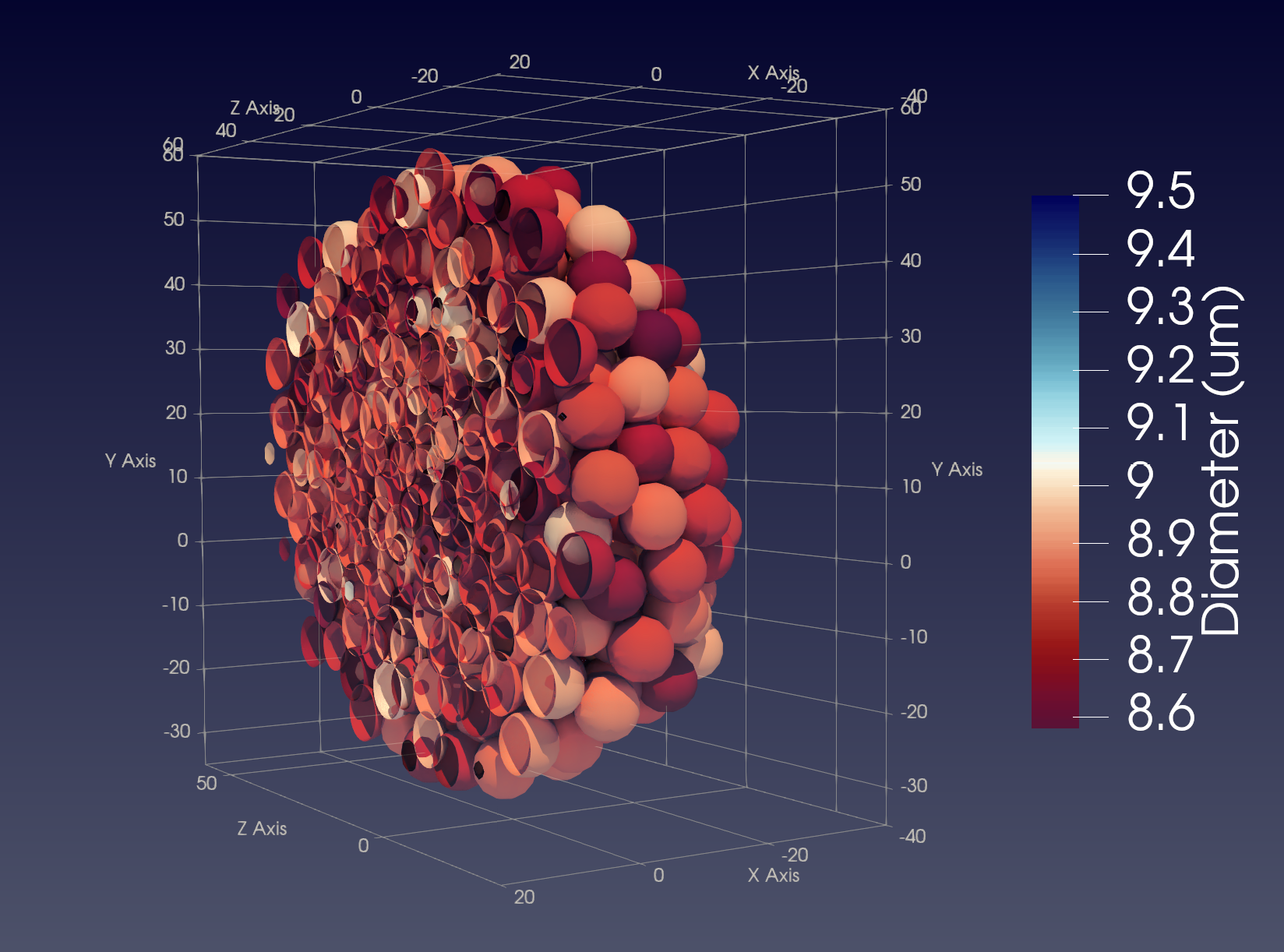}
    \caption{A visualization of the cell division module in BioDynaMo (cross-sectional view). The colors represent the diameter of the cells.}
    \label{fig:cell_division}
\end{figure}

 To quantify the impact of improving this operation for BioDynaMo, we run one of the available benchmarks that use all default operations (cell division module).
In this benchmark, a 3D grid of 262,144 cells of the same volume are spawned and proliferate for 10 iterations.
Once the cells are instantiated, in each iteration the same operations are executed: 1) cell proliferation, 2) neighborhood lookup, and 3) resolving the mechanical forces.
A visualization of cell proliferation in BioDynaMo with fewer cells and a longer runtime is shown in \Cref{fig:cell_division}. We profile this benchmark to get a better understanding of the computational bottlenecks in BioDynaMo.

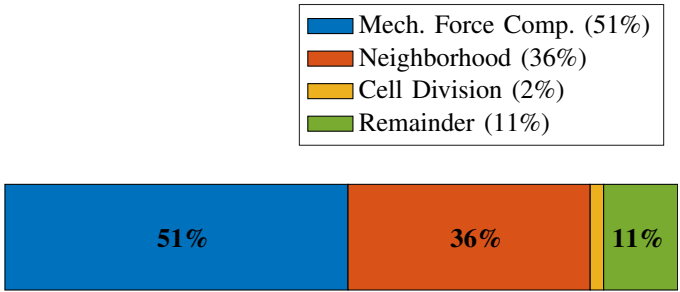
\begin{figure}[!t]
    \centering
%
%
\definecolor{mycolor1}{rgb}{0.00000,0.44700,0.74100}%
\definecolor{mycolor2}{rgb}{0.85000,0.32500,0.09800}%
\definecolor{mycolor3}{rgb}{0.92900,0.69400,0.12500}%
\definecolor{mycolor4}{rgb}{0.47,0.67,0.19}%
\begin{tikzpicture}

\begin{axis}[%
width=3.521in,
height=1.566in,
at={(0.758in,0.481in)},
scale only axis,
bar width=40,
xmin=0,
xmax=100,
ymin=0.7,
ymax=2.2,
axis line style={draw=none},
ticks=none,
legend style={legend cell align=left, align=left, draw=white!15!black}
]
\addplot[xbar stacked, fill=mycolor1, draw=black, area legend] table[row sep=crcr] {%
51	1\\
};
\addplot[forget plot, color=white!15!black] table[row sep=crcr] {%
0	-0.2\\
};
\addlegendentry{Mech. Force Comp. (51\%)}

\addplot[xbar stacked, fill=mycolor2, draw=black, area legend] table[row sep=crcr] {%
36	1\\
};
\addplot[forget plot, color=white!15!black] table[row sep=crcr] {%
0	-0.2\\
};
\addlegendentry{Neighborhood (36\%)}

\addplot[xbar stacked, fill=mycolor3, draw=black, area legend] table[row sep=crcr] {%
2	1\\
};
\addplot[forget plot, color=white!15!black] table[row sep=crcr] {%
0	-0.2\\
};
\addlegendentry{Cell Division (2\%)}

\addplot[xbar stacked, fill=mycolor4, draw=black, area legend] table[row sep=crcr] {%
11	1\\
};
\addplot[forget plot, color=white!15!black] table[row sep=crcr] {%
0	-0.2\\
};

\addlegendentry{Remainder (11\%)}

\end{axis}

\node[] at (4.3, 2) {\textbf{51\%}};
\node[] at (8.2, 2) {\textbf{36\%}};
\node[] at (10.35, 2) {\textbf{11\%}};

\end{tikzpicture}%
    \caption{Runtime profile of the cell division benchmark in BioDynaMo.}
    \label{fig:flamegraph}
\end{figure}


\par From \Cref{fig:flamegraph} we observe that the mechanical interactions operation (highlighted in blue) is the most time-consuming in the benchmark by a large margin.
Since this operation requires iterating over all agents, and in turn over all of their neighboring agents, this observation matches our prior expectation.
51\% of the benchmark's runtime is spent on the mechanical force calculations as described in \eqref{eq:default_force}, and 36\% is spent on updating the neighborhood list of each agent.
Updating the neighborhood is executed in two steps: 1) building a kd-tree, and 2) searching all the agents' neighbors within a specified radius.
\par A kd-tree is one of the many methods that can be used for a radial neighborhood search.
Considering that we want to offload this mechanical interactions operation to GPU, a more appealing method could be a uniform grid method.
The uniform grid method allows us to apply different techniques to improve the GPU version of the mechanical interactions operation, which we will discuss in this paper.

\section{Methodology} \label{sec:methodology}
\par In this section we will go over the implementation of the various improvements that were made on the existing mechanical interactions operation in BioDynaMo.
We use BioDynaMo v0.0.9-8b3d6c7 as the baseline version, which allows us to benefit more from GPU acceleration than the latest version presented in \cite{breitwieser2021biodynamo}, as the data are stored in a structs-of-arrays format, rather than arrays-of-structs.

\subsection{Uniform Grid Method}

\begin{figure}[!t]
    \centering
    \includegraphics[width=.2\textwidth]{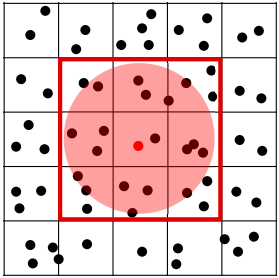}
    \caption{Finding the neighborhood of an agent using the uniform grid method. Displayed in 2D for simplicity.}
    \label{fig:regular_grid}
\end{figure}

\begin{figure}[!hbt]
    \centering
    \includegraphics[width=.48\textwidth]{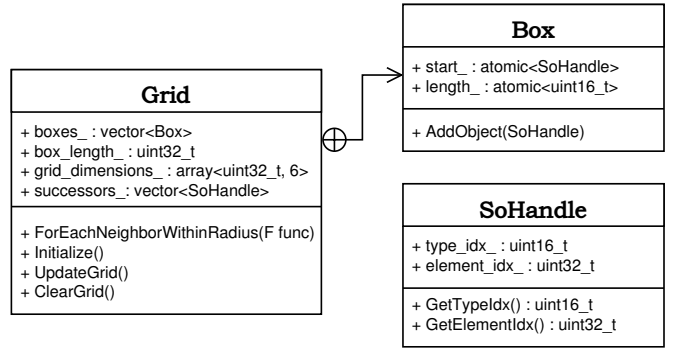}
    \caption{UML diagram of the class created for the uniform grid method.}
    \label{fig:sd_uml}
\end{figure}

The uniform grid method imposes a regularly-spaced 3D grid within the simulation space.
Each voxel of the grid contains only the agents that are confined within its subspace.
Finding the neighboring agents of a particular agent can be done by only taking into account the voxels surrounding that particular agent, as illustrated in 2D in \Cref{fig:regular_grid}.
The agent that we want to find the neighborhood for is colored red, and its interaction radius is highlighted in red.
We only consider the agents in the 9 surrounding voxels (27 in 3D) around which a red line is drawn in the figure.
We implement the uniform grid approach in BioDynaMo as a C++ class as illustrated in \Cref{fig:sd_uml} as a UML diagram.
For every simulation timestep, we reconstruct the uniform grid to take into account the addition, deletion, and movement of agents.
Each voxel (i.e. \texttt{Box}) keeps track of the number of agents it contains and the last object that was added.
Through the use of a linked list (\texttt{Grid::successors\_}) we can iterate through all objects inside a single \texttt{Box}.
The exact implementation details can be found in our Github repository\footnote{https://github.com/Senui/biodynamo/tree/paper-floats}.

\subsection{GPU Implementation}
\par We implement the uniform grid solution on the GPU using both CUDA and OpenCL to target GPUs from all major vendors.
To minimize the amount of CPU and GPU context switches, we decided to port the uniform grid algorithm as well as the mechanical force computation as a single GPU kernel.
Each GPU thread handles the mechanical interaction of one cell by 1) finding the cell's neighborhood, and 2) computing the mechanical forces between the cell and all the cells in its neighborhood.
The state data of all the agents in BioDynaMo are stored as structs-of-arrays (e.g. the position data of all agents are store contiguously in memory).
This allows us to copy the required state data for the mechanical interaction operation from the host DRAM to the GPU DRAM without first having to coalesce the data for all agents.

\subsection{Improvement I: Reduction in Floating-Point Precision}
\label{ssec:gpu_improvement-2}
\par BioDynaMo uses double-precision floating points (FP64) data types for all its floating-point data.
However, most consumer GPUs perform stronger in single-precision floating-point (FP32) operations.
This is a manifestation of the fact that GPU vendors primarily target the gaming industry and the field of artificial intelligence.
Game engines and machine learning frameworks rely mostly on single-precision floating-point operations, so GPU manufacturers designed their consumer GPUs with more FP32 logic units than their double-precision counterparts.
Some GPU vendors have dedicated cards for high-performance scientific computing that offer more FP64 logic units.
For agent-based simulations, other factors, such as choosing the correct runtime parameters for a model (e.g. initial agent attribute values, number of simulation steps, etc.), generally far outweigh the accuracy of the final results in comparison to the imprecision that could come forth from reducing the floating-point precision from double to single.
BioDynaMo has an extensive set of unit tests and integration tests that we can use to verify whether or not the reduction to FP32 affects the results.
Moreover, FP32 data types are half the size of FP64 data types in memory, which reduces the size of the buffers that need to be copied back and forth from the host to the device, leading to a potentially significant increase in throughput, and thus performance.

\subsection{Improvement II: Space-filling Curve Sorting}

\begin{figure}[!t]
    \centering
    \includegraphics[width=.25\textwidth]{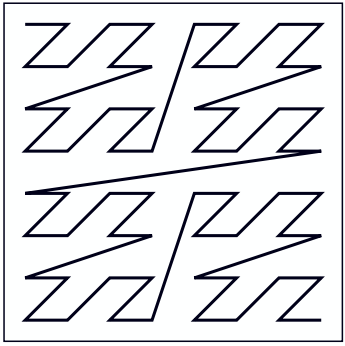}
    \caption{The path of a Z-order curve in 2D. Adapted from \cite{wikizorder}.}
    \label{fig:z_order_curve}
\end{figure}

\par CUDA and OpenCL organize threads in groups of threads; called blocks and workgroups, respectively.
The execution of the threads on the actual hardware is done in warps (generally in groups of 32 threads), with each warp executing the same instruction, but on different data (i.e. SIMT execution model).
BioDynaMo lays down the agents' data in memory in the order that the C++ objects were instantiated.
Each thread requires the data of the neighborhood of the simulation object it processes, which is not contiguous in memory, but rather scattered.
Consequently, each thread performs numerous scattered memory accesses, which will in most cases end up fetching the data from DRAM, which can degrade the performance significantly.
This could have been prevented if the data of agents that are close to each other in space are also laid down close to each other in memory.
This is where space-filling curves come in; more specifically the Z-order curve \cite{morton1966computer}.
A space-filling curve describes a path in multidimensional space that passes through the data points in consecutively local order, as illustrated in \Cref{fig:z_order_curve}.
A function that implements a space-filling curve can map multidimensional data (such as 3D Cartesian coordinates) to a one-dimensional array, where consecutive elements of that array are spatially local to each other.
For a Z-order curve, the \textit{Z-value} of each data point can be computed by binary interleaving its coordinate values and represents the index of the resulting one-dimensional array.
With regards to BioDynaMo, this would imply calculating the Z-values of all the agents and sorting their state data accordingly.
We anticipate that the cache line for accessing an agent will also contain the data of the agents in its neighborhood, and therefore reduces the number of fetches to DRAM.
A reduced number of fetches to DRAM should lead to a less data-starved execution pipeline, and therefore a higher throughput, and thus a reduction in the execution time for each simulation step.

\subsection{Improvement III: Using Shared Memory}
\label{ssec:gpu_improvement4}

\begin{figure}[t]
    \begin{subfigure}[t]{0.15\textwidth}
    \includegraphics[width=\hsize]{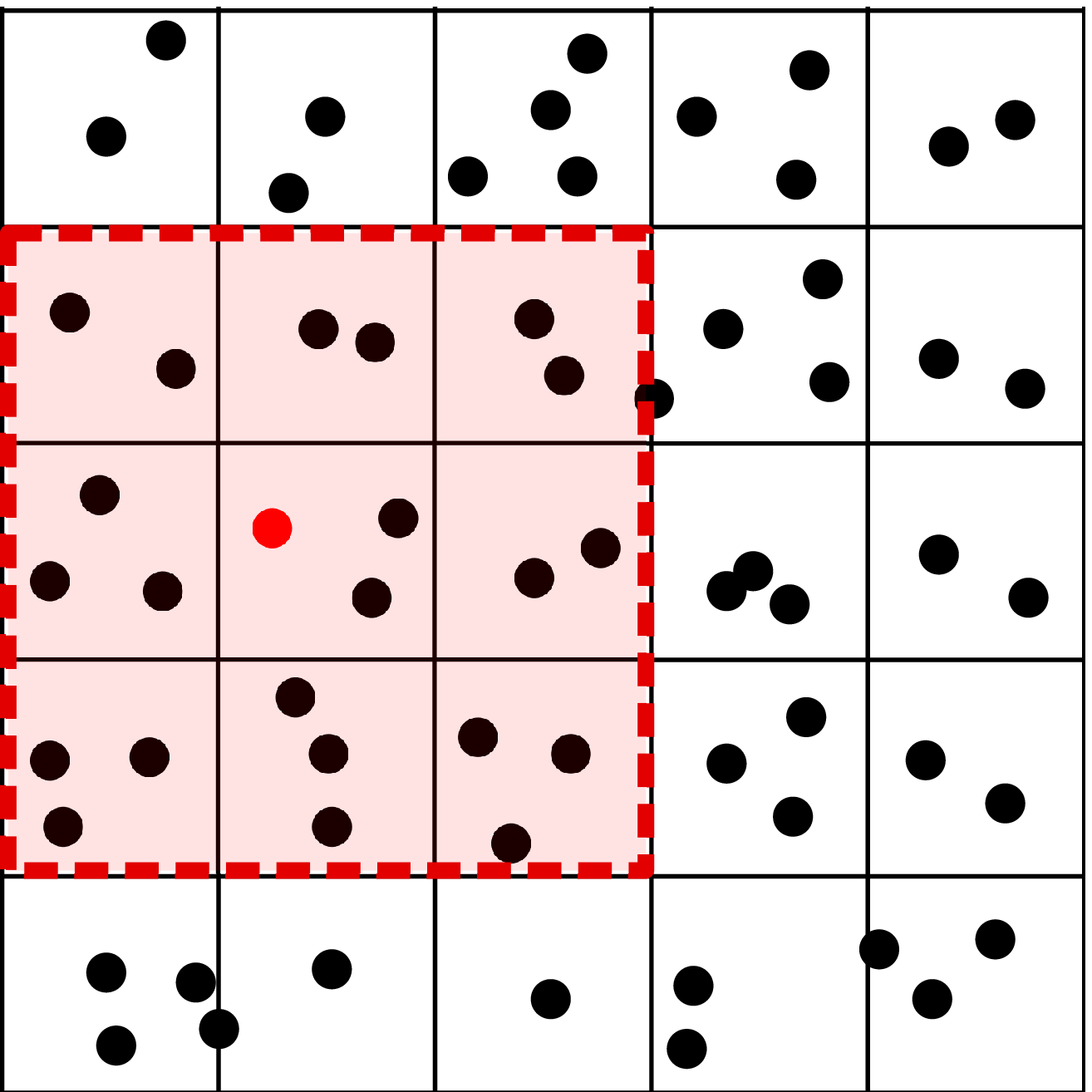}
    \caption{Data required by thread X.}
    \end{subfigure}
\hfill
    \begin{subfigure}[t]{0.15\textwidth}
    \includegraphics[width=\hsize]{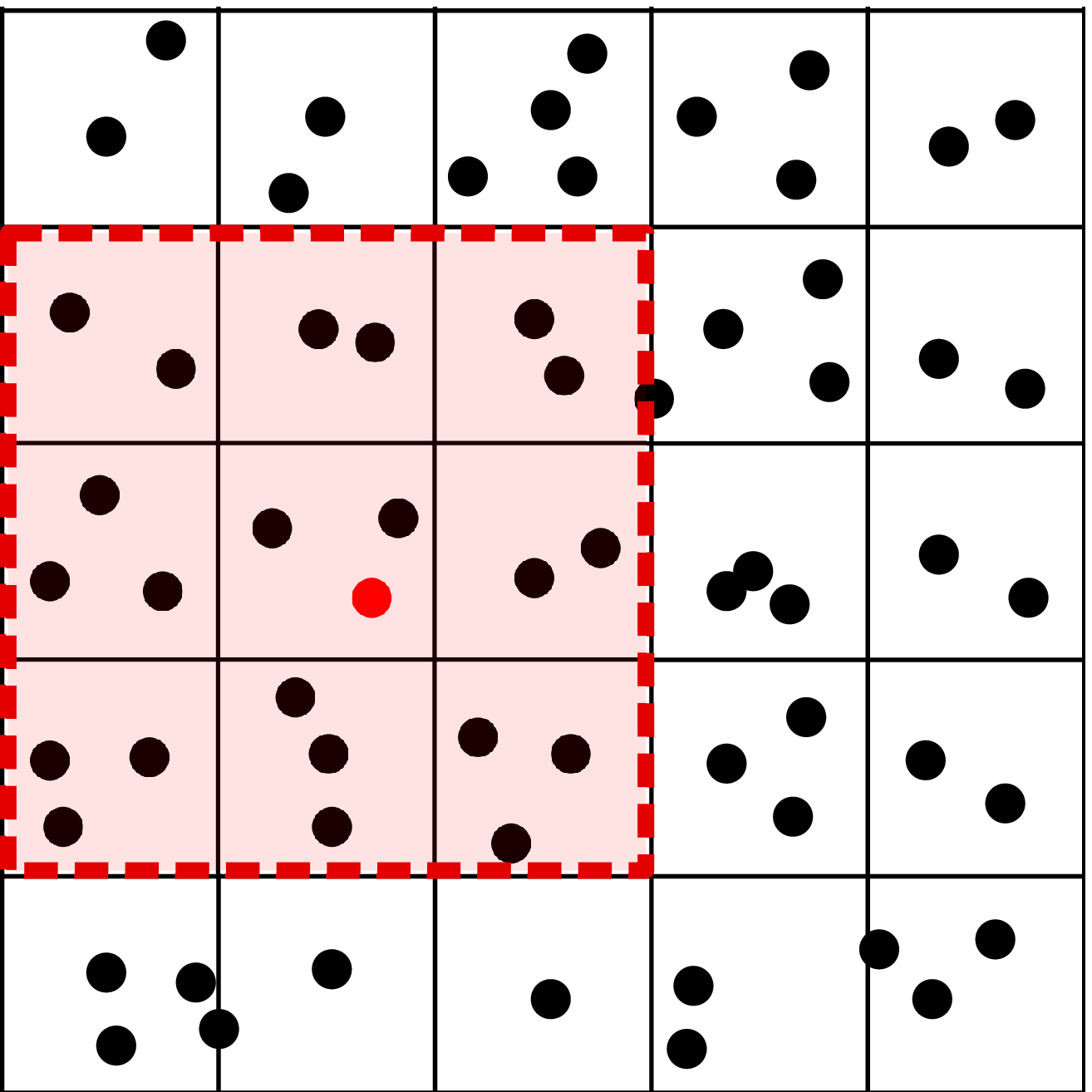}
    \caption{Data required by thread Y.}
\end{subfigure}
\hfill
    \begin{subfigure}[t]{0.15\textwidth}
    \includegraphics[width=\hsize]{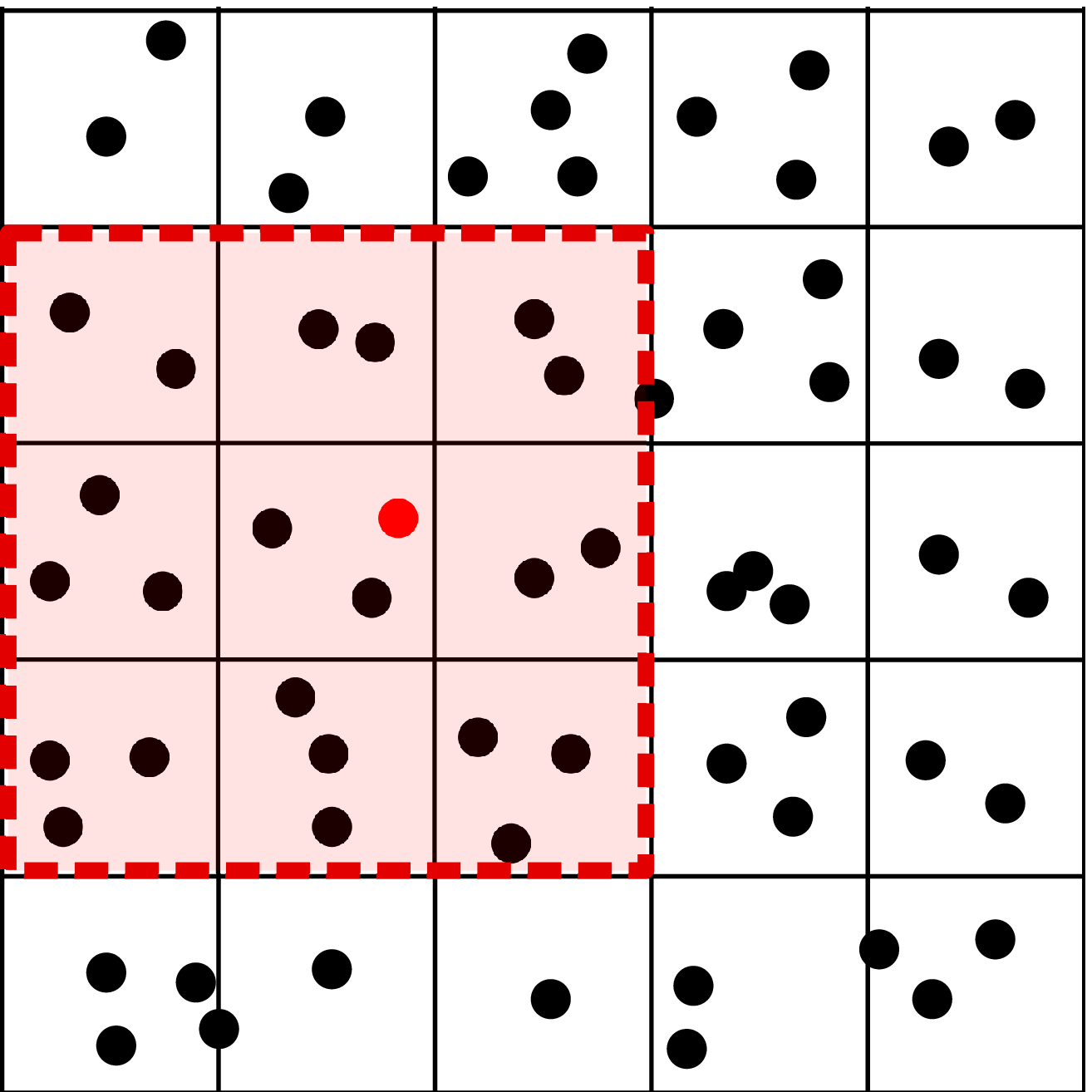}
    \caption{Data required by thread Z.}
\end{subfigure}
    \caption{Exploiting the reuse of neighboring simulation object data for the usage of shared memory resources on GPU.}
    \label{fig:gpu_improvement4}
\end{figure}

\par Most GPUs feature different types of on-chip memory, such as texture memory or shared memory.
In certain cases, storing data on on-chip memory drastically reduces the latency for fetching data during a GPU kernel execution, and could therefore improve the overall performance. 
In BioDynaMo, the concept of letting each GPU thread handle the mechanical interactions of one agent leaves little room for the shared memory resources of GPUs to be used.
The reason is that there is no reuse of data for threads within the same CUDA block (or OpenCL workgroup).
The kernel parallelizes the for loop over all agents, so each thread works on data that are independent of the threads in the same block.
To make use of shared memory, we need to create a kernel that allows multiple threads to work on mostly the same data.
It is here where we can reap the benefits of the uniform grid method that we implemented as an alternative to the kd-tree method.
We can exploit the fact that cells in the same voxel of the UG grid share the same neighboring voxels, and thus share the same simulation object candidates for their neighborhood.
Instead of parallelizing the for loop over all cells, we consider a kernel that would parallelize a loop over all voxels.
The threads that process the agents of a single voxel will need to reuse the neighborhood data, which can be stored in shared memory for low-latency memory fetches.
The concept is illustrated in \Cref{fig:gpu_improvement4}.
All the state data belonging to the agents that are within the highlighted region in \Cref{fig:gpu_improvement4} are stored in shared memory.
The shared memory objects are built in parallel by appending state data from agents of multiple voxels within the highlighted region.
To avoid race conditions, the use of atomic operations is required in building the shared memory objects in parallel.

\begin{table*}[!t]
\caption{Specifications of the systems used for benchmarking}\label{tab:gpu-hardware}
\begin{center}
\begin{tabular}{|c|c|c|c|c|c|c|c|c|}
\hline
&\textbf{GPU chip} & \textbf{GPU RAM} & \makecell{\textbf{Memory} \\ \textbf{bandwidth}} & \makecell{\textbf{Single-precision} \\ \textbf{performance}} & \makecell{\textbf{Double-precision} \\ \textbf{performance}} & \textbf{CPU chip} & \textbf{CPU cores} & \textbf{CPU DRAM} \\
\hline
\textbf{System A} & Nvidia GTX1080 Ti & 11GB & 484 GB/s & 11.34 TFLOPS & 0.354 TFLOPS & \makecell{Intel Xeon\\ E5-2640 v4} & \makecell{20 (2 sockets, \\40 threads)} & 256GB \\
\hline
\textbf{System B} & Nvidia Tesla V100 & 32GB & 900 GB/s & 15.7 TFLOPS & 7.8 TFLOPS & \makecell{Intel Xeon\\ Gold 6130} & \makecell{32 (2 sockets, \\64 threads)} & 187GB \\
\hline
\end{tabular}
\label{tab1}
\end{center}
\end{table*}

\section{Experimental Setup} \label{sec:experimental_setup}

\begin{figure}[!t]
    \centering
%
%
\definecolor{mycolor1}{rgb}{0.00000,0.44700,0.74100}%
\begin{tikzpicture}

\begin{axis}[%
width=0.25\textwidth,
height=1.566in,
at={(1.637in,0.481in)},
scale only axis,
bar shift auto,
log origin=infty,
xmode=log,
xmin=100,
xmax=200000, 
xminorticks=true,
xlabel style={font=\color{white!15!black}},
xlabel={Runtime (ms)},
ymin=-0.2,
ymax=9.2,
ytick={1,2,3,4,5,6,7,8},
yticklabels={{GPU Version III},{GPU Version II},{GPU Version I},{GPU Version 0},{  UG-method (20 threads)},{UG-method (serial)},{Baseline (20 threads)},{Baseline (serial)}},
axis background/.style={fill=white},
axis x line*=bottom,
axis y line*=left
]
\addplot[xbar, bar width=10, fill=mycolor1, draw=black, area legend] table[row sep=crcr] {%
274	1\\
199	2\\
527	3\\
1039	4\\
1910	5\\
14497	6\\
8226	7\\
25817	8\\
};
\node[right, align=left]
at (axis cs:274,1) {  274};
\node[right, align=left]
at (axis cs:199,2) {  199};
\node[right, align=left]
at (axis cs:527,3) {  527};
\node[right, align=left]
at (axis cs:1039,4) { 1039};
\node[right, align=left]
at (axis cs:1910,5) { 1910};
\node[right, align=left]
at (axis cs:14497,6) {14497};
\node[right, align=left]
at (axis cs:8226,7) { 8226};
\node[right, align=left]
at (axis cs:25817,8) {25817};
\end{axis}
\end{tikzpicture}%
    \caption{The runtime for various implementations of the mechanical interaction operation running benchmark A. The GPU results are obtained from the CUDA runtime on system A.}
    \label{fig:benchmarkA_runtimes}
\end{figure}
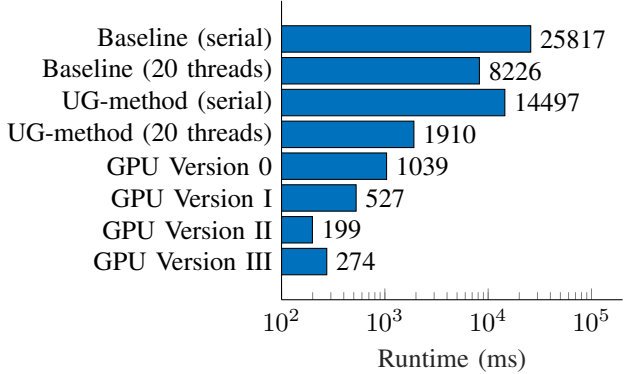

\par The hardware on which the evaluations are done belong to the CERN IT department and are tabulated in \Cref{tab:gpu-hardware}.
The CPUs of both systems consist of two physical sockets organized in a non-uniform memory access (NUMA) design.
To mitigate cross-NUMA effects on some of the benchmark results, we run those benchmarks on only one socket of the NUMA domains.
In practice, this was achieved by using the Linux utility tool \texttt{taskset}.
In \Cref{sec:results}, we explicitly mention the benchmarks that were run on a single NUMA domain.
The implementations and benchmarks can be found on Github\footnote{https://github.com/Senui/hicomb\_benchmarks}.
\par To profile the GPU kernel and the performance metrics we made use of \texttt{nvprof}, which is part of the CUDA SDK Toolkit.
Prior to recording the timing data for profiling GPU benchmarks, we run five iterations of the kernel to warm up the GPU.
This measure is necessary for the following reasons: 1) the GPU could initially be in a power-saving state and therefore not perform optimally on the first run, 2) just-in-time compilation of the kernel requires more time on the first compilation, 3) additional time could be taken for transferring the kernel binary to GPU memory.
\par To quantify the performance of our solutions, we perform three types of analyses.
First, we run the cell division benchmark (benchmark A) that was introduced in \Cref{sec:problem_definition}.
With this benchmark, we will quantify the performance of each solution in \Cref{sec:methodology}.
Second, we created a benchmark (benchmark B) to analyze the performance among models with different local neighborhood densities.
The cell division benchmark has a fixed average number of neighboring agents per agent, and therefore only represents models with the same neighborhood density.
With the second benchmark, we vary the average neighborhood density by spawning two million agents on random positions in variable-sized simulation space.
Consequently, the average number of neighboring agents per agent will be greater if the simulation space is smaller.
To maintain a constant neighborhood density over the simulated time, we set the maximum displacement value of each agent to zero.
The  (neighboring) agents will stay locked in space, and therefore the neighborhood density will stay constant.
The timing results of the benchmarks will exclude the model initialization time (creating the agents, assigning behaviors, etc.), and focus on the simulation performance.
Thirdly, to understand the performance limitations of the current GPU implementation, we perform a roofline analysis \cite{williams2009roofline} on the best performing GPU implementation.
Through this analysis, we will understand how far the current implementation is from the maximum attainable performance on system B.
We use the Empirical Roofline Tool (ERT) \cite{yang2018empirical} to measure the empirical performance numbers of system B and to generate the roofline analysis plot.
We retrieve the performance result (in GFLOP/s) and the arithmetic intensity (FLOPs/byte) of the GPU kernel with the use of \texttt{nvprof}.


\section{Results} \label{sec:results}

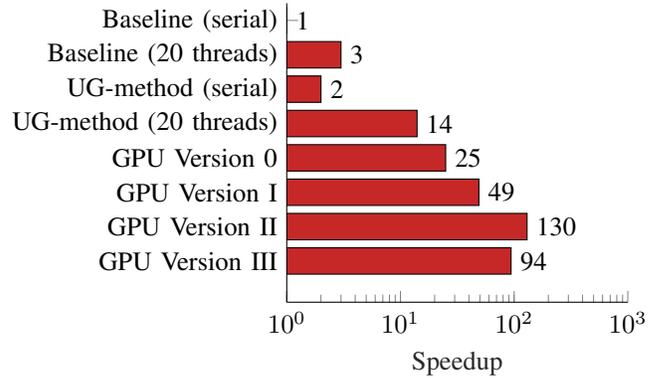
\begin{figure}[!t]
    \centering
%
%
\definecolor{mycolor1}{rgb}{0.77600,0.15600,0.15600}%
\begin{tikzpicture}

\begin{axis}[%
width=0.25\textwidth,
height=1.566in,
at={(1.637in,0.481in)},
scale only axis,
bar shift auto,
log origin=infty,
xmode=log,
xmin=1,
xmax=1000,
xminorticks=true,
xlabel style={font=\color{white!15!black}},
xlabel={Speedup},
ymin=-0.2,
ymax=8.5,
ytick={1,2,3,4,5,6,7,8},
yticklabels={{GPU Version III},{GPU Version II},{GPU Version I},{GPU Version 0},{  UG-method (20 threads)},{UG-method (serial)},{Baseline (20 threads)}, {Baseline (serial)}},
axis background/.style={fill=white},
axis x line*=bottom,
axis y line*=left
]
\addplot[xbar, bar width=10, fill=mycolor1, draw=black, area legend] table[row sep=crcr] {%
94	1\\
130	2\\
49	3\\
25	4\\
14	5\\
2	6\\
3	7\\
1	8\\
};
\node[right, align=left]
at (axis cs:94,1) { 94};
\node[right, align=left]
at (axis cs:130,2) {130};
\node[right, align=left]
at (axis cs:49,3) { 49};
\node[right, align=left]
at (axis cs:25,4) { 25};
\node[right, align=left]
at (axis cs:14,5) { 14};
\node[right, align=left]
at (axis cs:2,6) {  2};
\node[right, align=left]
at (axis cs:3,7) {  3};
\node[right, align=left]
at (axis cs:1,8) {  1};
\end{axis}
\end{tikzpicture}%
    \caption{The speedup with respect to the serial baseline version as obtained with benchmark A. The GPU results are obtained from the CUDA runtime on system A.}
    \label{fig:benchmarkA_speedups}
\end{figure}

\begin{figure}[!t]
    \centering
%
%
\definecolor{mycolor1}{rgb}{0.00000,0.44700,0.74100}%
\definecolor{mycolor2}{rgb}{0.85000,0.32500,0.09800}%
\definecolor{mycolor3}{rgb}{0.92900,0.69400,0.12500}%
\definecolor{mycolor4}{rgb}{0.49400,0.18400,0.55600}%
\definecolor{mycolor5}{rgb}{0.46600,0.67400,0.18800}%
\definecolor{mycolor6}{rgb}{0.30100,0.74500,0.93300}%
\def\barWidth{2}
\begin{tikzpicture}

\begin{axis}[%
width=0.4\textwidth,
height=2.5in,
at={(0.758in,0.481in)},
scale only axis,
bar shift auto,
log origin=infty,
xmin=0.493333333333333,
xmax=8.50666666666667,
xtick={1,2,3,4,5,6,7,8},
xticklabels={{1},{3},{6},{11},{17},{27},{35},{47}},
xlabel style={font=\color{white!15!black}},
xlabel={Number of neighbors per agent},
ymode=log,
ymin=1000,
ymax=9000000,
yminorticks=true,
ylabel style={font=\color{white!15!black}},
ylabel={Runtime (ms)},
axis background/.style={fill=white},
xmajorgrids,
ymajorgrids,
yminorgrids,
legend style={at={(0.03,0.97)}, {nodes={scale=0.6, transform shape}}, anchor=north west, legend cell align=left, align=left, draw=white!15!black}
]
\addplot[ybar, bar width=\barWidth, fill=mycolor1, draw=black, area legend] table[row sep=crcr] {%
1	253979.6\\
2	308048.8\\
3	350276.4\\
4	417854.8\\
5	466308\\
6	559292.4\\
7	629726.4\\
8	729408.4\\
};
\addlegendentry{Intel Xeon 6130 (4 threads)}

\addplot[ybar, bar width=\barWidth, fill=mycolor2, draw=black, area legend] table[row sep=crcr] {%
1	218357.6\\
2	242994\\
3	257788.6\\
4	291769.6\\
5	330370.6\\
6	382151.6\\
7	414960.6\\
8	463514.2\\
};
\addlegendentry{Intel Xeon 6130 (8 threads)}

\addplot[ybar, bar width=\barWidth, fill=mycolor3, draw=black, area legend] table[row sep=crcr] {%
1	197488\\
2	210015.6\\
3	221753.4\\
4	244417.4\\
5	258123.4\\
6	284217.2\\
7	305370\\
8	329436.4\\
};
\addlegendentry{Intel Xeon 6130 (16 threads)}

\addplot[ybar, bar width=\barWidth, fill=mycolor4, draw=black, area legend] table[row sep=crcr] {%
1	178317.6\\
2	186628.6\\
3	193826.6\\
4	203811.4\\
5	215403.4\\
6	228007.6\\
7	239248\\
8	256171.6\\
};
\addlegendentry{Intel Xeon 6130 (32 threads)}

\addplot[ybar, bar width=\barWidth, fill=mycolor5, draw=black, area legend] table[row sep=crcr] {%
1	179015.6\\
2	182828.6\\
3	190092.8\\
4	197905.6\\
5	201375\\
6	210196\\
7	218308\\
8	226664.8\\
};
\addlegendentry{Intel Xeon 6130 (64 threads)}

\addplot[ybar, bar width=\barWidth, fill=mycolor6, draw=black, area legend] table[row sep=crcr] {%
1	1584\\
2	1623.8\\
3	1663.2\\
4	1997\\
5	2134\\
6	2479.4\\
7	2714\\
8	3175.4\\
};
\addlegendentry{Tesla V100}

\end{axis}
\end{tikzpicture}%
    \caption{The runtime of benchmark B for a varying neighborhood density. The Intel Xeon entries represent the baseline version. The Tesla V100 entries represent the best performing GPU implementation. The GPU results are obtained from the CUDA runtime on system B.}
    \label{fig:benchmarkB_runtimes}
\end{figure}
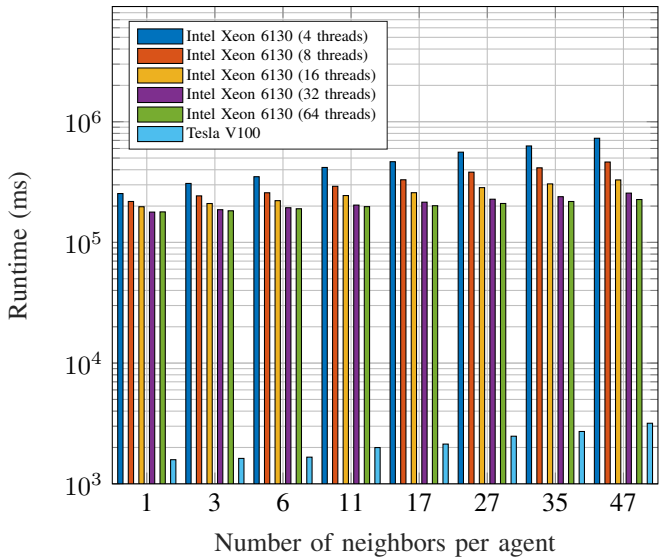

\par \Cref{fig:benchmarkA_runtimes} shows the runtimes obtained from running benchmark A for the various implementations of the mechanical interaction operation in BioDynaMo.
\Cref{fig:benchmarkA_speedups} shows the obtained speedups comparison to the serial baseline version.
Note that the x-axis is scaled logarithmically in both figures.
The order of the bar charts follows from the order in which the versions were introduced in \Cref{sec:methodology}.
Consecutive GPU versions include the implementation of the prior version, so for example GPU version II includes the changes made for GPU version I.
The results in \Cref{fig:benchmarkA_speedups} are obtained from running benchmark A on system A.

\par The serial uniform grid (UG) method performs twice as fast as the serial kd-tree method.
On all 20 cores of the system (on a single NUMA domain), the UG method is $\frac{8226}{1910} = 4.3$ times faster than the kd-tree method.
This can be attributed to the parallel construction of the uniform grid as opposed to the serial construction of the kd-tree.

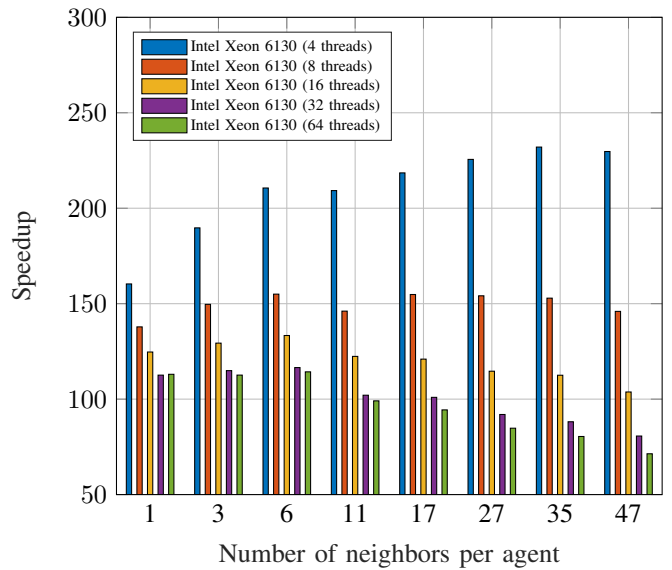
\begin{figure}[!t]
    \centering
%
%
\definecolor{mycolor1}{rgb}{0.00000,0.44700,0.74100}%
\definecolor{mycolor2}{rgb}{0.85000,0.32500,0.09800}%
\definecolor{mycolor3}{rgb}{0.92900,0.69400,0.12500}%
\definecolor{mycolor4}{rgb}{0.49400,0.18400,0.55600}%
\definecolor{mycolor5}{rgb}{0.46600,0.67400,0.18800}%
\def\barWidth{2}
\begin{tikzpicture}

\begin{axis}[%
width=0.4\textwidth,
height=2.5in,
at={(0.758in,0.481in)},
scale only axis,
bar shift auto,
xmin=0.507692307692308,
xmax=8.49230769230769,
xtick={1,2,3,4,5,6,7,8},
xticklabels={{1},{3},{6},{11},{17},{27},{35},{47}},
xlabel style={font=\color{white!15!black}},
xlabel={Number of neighbors per agent},
ymin=50,
ymax=300,
ylabel style={font=\color{white!15!black}},
ylabel={Speedup},
axis background/.style={fill=white},
xmajorgrids,
ymajorgrids,
legend style={at={(0.03,0.97)}, {nodes={scale=0.6, transform shape}}, anchor=north west, legend cell align=left, align=left, draw=white!15!black}
]
\addplot[ybar, bar width=\barWidth, fill=mycolor1, draw=black, area legend] table[row sep=crcr] {%
1	160.34\\
2	189.71\\
3	210.6\\
4	209.24\\
5	218.51\\
6	225.58\\
7	232.03\\
8	229.71\\
};
\addplot[forget plot, color=white!15!black] table[row sep=crcr] {%
0.507692307692308	0\\
8.49230769230769	0\\
};
\addlegendentry{Intel Xeon 6130 (4 threads)}

\addplot[ybar, bar width=\barWidth, fill=mycolor2, draw=black, area legend] table[row sep=crcr] {%
1	137.85\\
2	149.65\\
3	155\\
4	146.1\\
5	154.81\\
6	154.13\\
7	152.9\\
8	145.97\\
};
\addplot[forget plot, color=white!15!black] table[row sep=crcr] {%
0.507692307692308	0\\
8.49230769230769	0\\
};
\addlegendentry{Intel Xeon 6130 (8 threads)}

\addplot[ybar, bar width=\barWidth, fill=mycolor3, draw=black, area legend] table[row sep=crcr] {%
1	124.68\\
2	129.34\\
3	133.33\\
4	122.39\\
5	120.96\\
6	114.63\\
7	112.52\\
8	103.75\\
};
\addplot[forget plot, color=white!15!black] table[row sep=crcr] {%
0.507692307692308	0\\
8.49230769230769	0\\
};
\addlegendentry{Intel Xeon 6130 (16 threads)}

\addplot[ybar, bar width=\barWidth, fill=mycolor4, draw=black, area legend] table[row sep=crcr] {%
1	112.5742\\
2	114.9332\\
3	116.5384\\
4	102.0588\\
5	100.9388\\
6	91.9608\\
7	88.1533\\
8	80.6738\\
};
\addplot[forget plot, color=white!15!black] table[row sep=crcr] {%
0.507692307692308	0\\
8.49230769230769	0\\
};
\addlegendentry{Intel Xeon 6130 (32 threads)}

\addplot[ybar, bar width=\barWidth, fill=mycolor5, draw=black, area legend] table[row sep=crcr] {%
1	113.0149\\
2	112.5931\\
3	114.2934\\
4	99.1015\\
5	94.365\\
6	84.777\\
7	80.4377\\
8	71.3815\\
};
\addplot[forget plot, color=white!15!black] table[row sep=crcr] {%
0.507692307692308	0\\
8.49230769230769	0\\
};
\addlegendentry{Intel Xeon 6130 (64 threads)}

\end{axis}
\end{tikzpicture}%
    \caption{The speedups with respect to the baseline version (for various numbers of threads) as obtained with benchmark B for a varying neighborhood density. The GPU results are obtained from the CUDA runtime on system B.}
    \label{fig:benchmarkB_speedups}
\end{figure}

\par The initial version of the GPU implementation (GPU version 0 in \Cref{fig:benchmarkA_speedups}) of the UG method already offers an $\frac{8226}{1039} = 7.9\times$ speedup as compared to the multithreaded baseline version and is $\frac{1910}{1039} = 1.8$ times faster than its multithreaded CPU version.
Even though the kernel is not yet optimized, due to the massively parallel architecture of the GPU we are able to attain a significant speedup compared to the multithread CPU version.
\par From \Cref{fig:benchmarkA_speedups} we can see about a $\frac{1039}{527} = 2.0$ speedup gained from reducing the data types that define a cell's state from doubles to floats.
From \Cref{tab:gpu-hardware} we can see that the FP32 throughput is 32 times greater than the FP64 throughput.
From our speedup result, it becomes clear that the current GPU solution is limited by the memory bandwidth.
Since FP32 data types are 4 bytes and FP64 data types are 8 bytes, the expected speedup of a GPU application that is memory bound and heavily relies on floating-point operations is two.
We verified that the correctness of the simulations was not affected as a result of reducing the floating-point precision by running the unit tests and integration tests that are included in the testing suite of BioDynaMo.
\par Sorting the agents' state data based on a space-filling curve proved to reduce the execution time significantly, namely $\frac{527}{199} = 2.6$ times in comparison to the previous GPU version.
This speedup confirms that the GPU kernel enjoys more spatial data locality when the agents' state data is sorted.
As a result, memory accesses are more coalesced, which in turn leads to an increase in cache hits.
This reduces the overall latency of obtaining the required neighborhood data from memory.
\par Redesigning the GPU kernel to utilize shared memory resources appears to worsen the overall performance by $28\%$.
One of the reasons we found that causes the kernel performance to deteriorate, is the introduction of atomic operations in the kernel.
The use of atomics is necessary to build the shared data structures that were introduced in \Cref{ssec:gpu_improvement4} in parallel.
However, this causes stalling when multiple threads try to update the same shared data object.
Moreover, the kernel needs to perform boundary checks on the blocks (CUDA) or workgroups (OpenCL) that are being executed by the GPU, which gives rise to thread divergence.
\par \Cref{fig:benchmarkB_runtimes} and \Cref{fig:benchmarkB_speedups} summarize the results from running benchmark B on system B.
The CPU results up to 32 threads were obtained by running on a single NUMA domain on system B.
\Cref{fig:benchmarkB_runtimes} shows the runtimes of the multithreaded baseline version (4, 8, 16, 32, and 64 threads) and the best performing GPU version (GPU version II), for a varying number of neighboring agents per agent.
From the figure, it becomes clear that increasing the number of threads in a CPU-only runtime only reduces the runtime marginally, whereas GPU co-processing shows a significant reduction in runtime. 
\Cref{fig:benchmarkB_speedups} shows the speedup of the GPU runtime in comparison to the multithreaded baseline version.
We observe that the speedup in comparison to the baseline version running with 4 threads lies between $160\times$ to $232\times$, depending on the neighborhood density.
For the baseline version with 64 threads, the speedup lies between $71\times$ to $113\times$.
These results imply that simulations that are densely populated, enjoy a speedup of up to two orders of magnitude when accelerating their workload with a GPU.
Simulations that would normally take days on a multi-core CPU can be completed in hours on systems that feature a GPU.
The significant reduction in simulation runtime allows researchers in the field of biological ABS to scale out their models and still obtain results rapidly.

\par In \Cref{fig:benchmarkB_speedups} we notice that the GPU performance gain stagnates, or even decreases, as the neighborhood density increases.
The GPU kernel parallelizes the mechanical interaction computation for all agents, but the loop over all neighboring agents is serial.
Consequently, this becomes the bottleneck for models with a high neighborhood density.
We would like to investigate this solution by exploring \textit{dynamic parallelism} \cite{jones2012introduction} in existing GPU programming models.
We hypothesize that parallelizing the serial loop over the neighborhood alleviates the bottleneck that is manifested in \Cref{fig:benchmarkB_speedups}.
\par
\begin{figure}[!t]
    \centering
    \includegraphics[width=.53\textwidth]{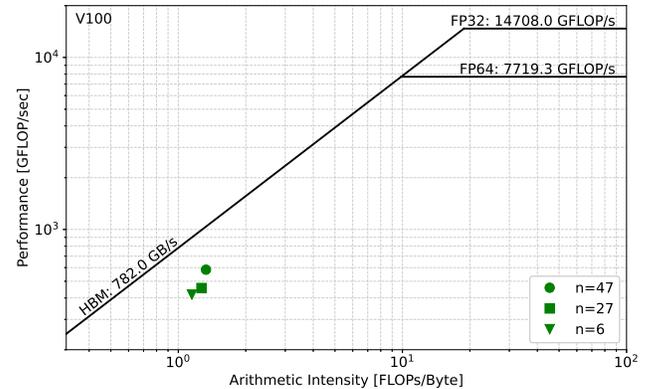}
    \caption{GPU roofline model analysis of various neighborhood densities on system B, where $n$ is the number of neighbors per agent.}
    \label{fig:roofline}
\end{figure}

 From the roofline model analysis in \Cref{fig:roofline}, we see that the best performing GPU implementation is still an order of magnitude away from the maximum attainable single-precision floating-point performance on system B.
The data points are however close to the roof that represents the upper bound of the device memory bandwidth (HBM), which indicates that the kernel is close to being memory-bound.
Future improvements to the kernel must focus on alleviating the strain on data transfer between the GPU and the GPU memory.
Investigating other caching methods to bypass the HBM bandwidth roofline should be the main priority for future improvements.
We observe that the kernel is able to attain higher performance with a higher neighborhood density.
Based on the percentage of L2 cache reads relative to the number of total (L2 + HBM) memory reads, as obtained by \texttt{nvprof}, we believe this to be the result of increased cache reuse of the neighborhood state data per agent.
For $n = 47$ this percentage is 41.3\%, for $n = 27$ it is 40.6\% and for $n = 6$ it is 39.4\%.

\section{Conclusion} \label{sec:conclusions}

\par The goal of this work was to perform a comparative study of the acceleration potential of GPU co-processing in BioDynaMo to enable fast simulation of large-scale and complex biological models.
To understand what the most effective way is to improve the performance of simulations such that large-scale and complex models can be implemented, we profiled the simulations that BioDynaMo is currently capable of running.
We discovered that the mechanical interactions operation was the computational bottleneck by a large margin, due to the required data of local neighboring agents.
We implemented a method alternative to the kd-tree method, the uniform grid (UG) method, which proved to be an excellent candidate for exploiting the parallel architecture of GPUs for performance gain.
Not only did the UG method outperform the kd-tree method on CPU, but it opened up possibilities to exploit the advantages that GPUs offer.
The final GPU kernel implementation resulted in speedups between $71\times$ to $232\times$ in comparison to the multithreaded baseline version, depending on the number of neighboring agents per agent and the number of threads the baseline is executed with.
This result enables researchers of cellular agent-based models to rapidly obtain biologically insightful simulations with BioDynaMo.

\section*{Funding}
This  work  was  supported  by  the  CERN Knowledge  Transfer  office  [to L.B.] and CERN openlab [to A.H.].

\printbibliography

\end{document}